# Synthesis of Monolayer Ice on a Hydrophobic Metal Surface

Qiaoxiao Zhao,[1,2,‡] Meiling Xu,[3,‡] Dong Li,[1,2] Zhicheng Gao,[1,2] Yudian Zhou,[1,2] Wenbo Liu,[1,2] Jingyan Chen,[3] Peng Cheng,[1,2] Sheng Meng,[1,2] Kehui Wu,[4] Yanchao Wang,[5*] Lan Chen,[1,2*] Baojie Feng,[1,2*]

1 Institute of Physics, Chinese Academy of Sciences, Beijing, 100190, China

2 School of Physical Sciences, University of Chinese Academy of Sciences, Beijing, 100049, China

3 Laboratory of Quantum Functional Materials Design and Application, School of Physics and Electronic Engineering, Jiangsu Normal University, Xuzhou 221116, China

4 Tsientang Institute for Advanced Study, Zhejiang 310024, China

5 Key Laboratory of Material Simulation Methods & Software of Ministry of Education, College of Physics, Jilin University, Changchun 130012, China

**ABSTRACT:** Understanding water–metal interactions is central to disciplines spanning catalysis, electrochemistry, and atmospheric science. Monolayer ice phases are well established on hydrophilic surfaces, where strong water–substrate interactions stabilize ordered hydrogen-bond networks. In contrast, their formation on hydrophobic metals has been deemed thermodynamically unfavourable, with water typically assembling into amorphous films, three-dimensional crystallites, or interlocked bilayer ice. Here, we demonstrate the synthesis of a monolayer ice phase on the hydrophobic Au(111) surface using a low-energy-electron–assisted growth method. Combined experimental characterizations including low-energy electron diffraction, angle-resolved photoemission spectroscopy, and X-ray photoelectron spectroscopy, complemented by first-principles calculations, prove that the monolayer ice phase composes of intact water molecules. This approach provides a generalizable strategy for stabilizing ordered two-dimensional ice on inert substrates and offers new insight into the interplay between water and low-energy electrons at hydrophobic interfaces.

## Introduction

Water on metal interfaces governs a wide range of physicochemical phenomena, including lubrication, heterogeneous catalysis, corrosion, and electrochemical reactions(1-3). In particular, the adsorption and organization of water molecules on metal surfaces have long served as model systems for investigating solid–liquid interactions at the molecular scale. The interfacial structures(4) emerge from a delicate balance between water–water hydrogen bonding and water–substrate interactions, which are often comparable in magnitude. This interplay gives rise to a rich spectrum of morphologies, from disordered clusters to two-dimensional (2D) ice crystalline phases, even on atomically flat metal surfaces(5-12). The microscopic understanding of interfacial water, especially its structural diversity and dynamic behavior, critically influences surface reactivity, wetting, and ion transport, with implications for electrocatalysis, nanofluidics, and atmospheric chemistry, but decades of experimental and theoretical studies remains contradiction(13).

Ordered 2D ice phases have been reported predominantly on hydrophilic metal substrates, where strong water–substrate interactions stabilize extended hydrogen-bond networks. On hydrophobic metals, by contrast, it is generally accepted that water preferentially forms three-dimensional ice crystallites, amorphous films, or other metastable aggregates owing to the weak adsorption energy(14,15). A notable exception is the case on Au(111)(12,16-21), which supports an interlocked bilayer ice phase: two planar hexagonal ice layers connected by a fully hydrogen-bonded network. Both real-space atomic images and theoretical calculations have shown that the interlocked bilayer can be stabilized even without strong substrate coupling, making it unusually robust among hydrophobic interfaces. In contrast, monolayer ice is widely regarded as unstable on hydrophobic metals due to its reduced hydrogen bonding capacity, and has so far been observed only on hydrophilic substrates where the strong water–substrate interactions play a crucial role (8,9,22,23).

Here, we report the synthesis of a monolayer ice phase on the hydrophobic Au(111) surface using a low-energy-electron–assisted (LEEA) growth method, as shown in Fig. 1a. By injecting low-energy electrons into interlocked bilayer ice, we induce a phase transformation to a previously unreported monolayer structure composed of intact $H_2O$ molecules without detectable dissociation. Low-energy electron diffraction (LEED), angle-resolved photoemission spectroscopy (ARPES), and X-ray photoelectron spectroscopy (XPS), combined with first-principles calculations, reveal the atomic structure, electronic properties, and stability of this phase. Our results show that electron injection can act as an effective stimulus to control the phase of 2D ice. This discovery expands the phase diagram of interfacial water and provides a platform for exploring quantum and collective phenomena in 2D light-element systems.

## Results and Discussion

### Synthesis of monolayer ice on Au(111)

In our experiments, we firstly prepared the well-established interlocked bilayer ice (12,16,17,24) on Au(111) via water vapor deposition at approximately 120 K (see Methods). As shown in the low-energy electron diffraction (LEED) patterns (Fig. 1b), this phase can be identified by a ($\sqrt{3}\times\sqrt{3}$)R30° superstructure with respect to the 1×1 lattice of Au(111), as shown in Fig. 1b. Thicker multilayer ice forms amorphous structures without long-range order, as shown in Supplementary Fig. S1. This growth behavior ensures that the ordered ($\sqrt{3}\times\sqrt{3}$)R30° phase corresponds to the interlocked bilayer ice. Upon increasing the electron beam energy and exposure time during LEED measurements, the ($\sqrt{3}\times\sqrt{3}$)R30° pattern progressively transformed into a new 2×2 superstructure (Fig. 1c). The electron energies required for the structural transformation range from several tens to a few hundred electron volts, and the transformation rate increases with higher electron flux, independent of the electron energy (Supplementary Fig. S2). These observations demonstrate that electron irradiation is the driving stimulus for the phase transition. Notably, the transformation is not reversible: re-dosing $H_2O$ onto the 2×2 phase does not restore the ($\sqrt{3}\times\sqrt{3}$)R30° phase (Supplementary Fig. S3).

Then the 2×2 LEED pattern completely disappeared when we annealed the sample to 150 K, leaving only the 1×1 diffraction spots corresponding to the clean Au(111) substrate (Supplementary Fig. S4d). The recovery of the clean Au(111) surface is further confirmed by our temperature-dependent XPS and ARPES measurements (Supplementary Fig. S5 and S6). The reappearance of the characteristic Shockley surface state indicates that the Au(111) substrate is not reconstructed during the entire process. This behavior suggests that the 2D ice desorbs thermally at elevated temperatures. The critical desorption temperature of the 2×2 phase is comparable to that of the interlocked bilayer ice and of other reported water phases on metal interfaces(25), implying that the phase is composed of intact $H_2O$ molecules. Significant dissociation into hydroxyl species is unlikely, as OH typically binds more strongly to metal surfaces and desorbs at substantially higher temperatures(26). Prolonged exposure to the electron beam after formation of the 2×2 phase also led to the disappearance of its LEED pattern, consistent with electron-stimulated desorption of water molecules.

### XPS study of monolayer ice on Au(111)

To probe the chemical nature of the newly observed 2×2 phase, we performed *in situ* X-ray photoelectron spectroscopy (XPS). The O 1s spectrum of thick amorphous ice deposited at ~70 K exhibits a peak at 533.7 eV (Fig. 2a). In contrast, both the interlocked bilayer ice and the 2×2 phase display a red shift of the O 1s peak to 533.0 eV (Fig. 2b). This shift likely reflects increased average charge transfer between the Au(111) and the 2D ice layers compared with bulk-like amorphous ice, and is consistent with prior reports for ultrathin ice on other metal surfaces, including 533.5 eV on Cu(111)(27,28) and 532.2 eV on Pt(111)(29-31).

The O 1s peak of the 2×2 phase shows no shift or splitting relative to the bilayer ice, indicating that low-energy electron irradiation does not induce water dissociation. This conclusion is consistent with earlier studies of water adsorption on noble metals such as Cu(111)(27), Ag(110)(32), Au(110)(33), and Au(111)(34), where intact $H_2O$ molecules persist under similar conditions. In contrast, LEEA growth of water on Ag(111)(26,35) has been shown to produce hydroxylated monolayer ice, whose O 1s peak exhibits a distinct shoulder from OH species. The absence of such features here confirms that the 2×2 phase on Au(111) is composed exclusively of intact $H_2O$ molecules.

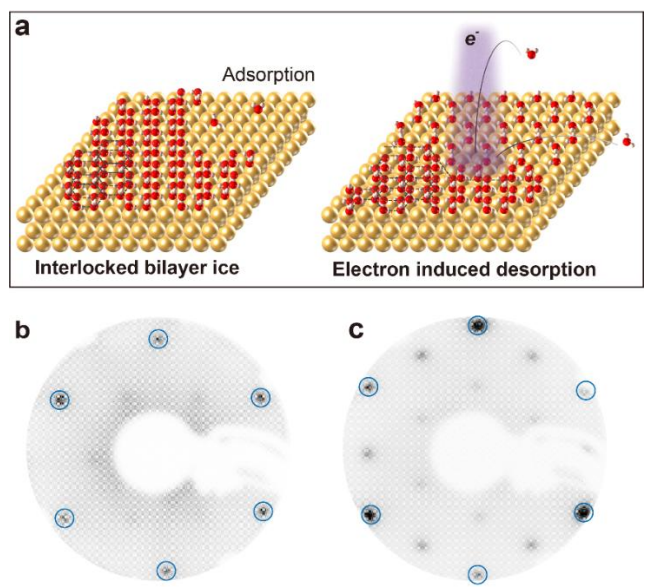

**Fig. 1.** Structure characterization of two types of 2D ice phase. (a) Schematic diagram of the electron beam-induced 2D $H_2O$ phase transition on Au(111). (b, c) LEED patterns of the ($\sqrt{3}\times\sqrt{3}$)R30° and the 2×2 ice phases on Au(111), respectively. Blue circles mark the diffraction spots of the Au(111) substrate.

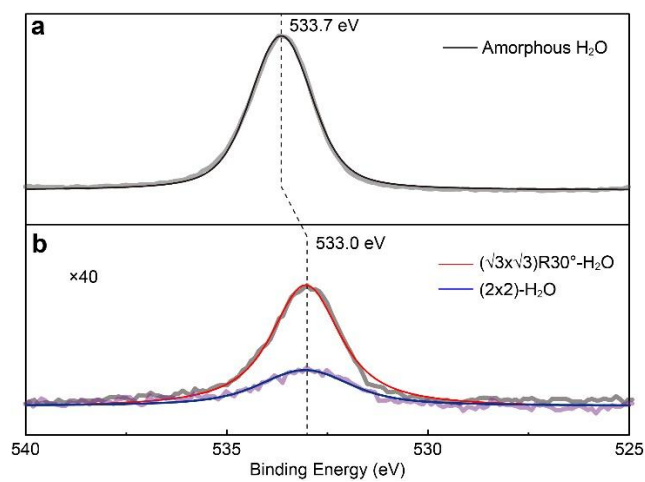

**Fig. 2.** XPS spectra of amorphous and ordered ice phases on Au(111). (a) O 1s core-level spectrum of amorphous $H_2O$ adsorbed on Au(111) at 70 K. (b) O 1s spectra of the interlocked bilayer ice phase, exhibiting the ($\sqrt{3}\times\sqrt{3}$) R30° superstructure (red), and the monolayer ice phase with the 2×2 superstructure (blue). The spectra were fitted using Voigt line shapes, accounting for the convolution of Gaussian instrumental broadening and Lorentzian lifetime broadening, with a Shirley background subtracted in each case. The uncertainty of the peak area ratio was determined using standard error-propagation rules.

Quantitative analysis of the integrated O 1s peak areas reveals a coverage ratio of 2.74±0.18 between the bilayer ice and the 2×2 phase. This substantial decrease in oxygen signal indicates that electron irradiation induces significant desorption of $H_2O$ molecules from the Au(111) surface. The reduced O 1s intensity suggest that the 2×2 phase has a monolayer ice structure, which contains fewer water molecules per unit area than the bilayer ice.

Structure model of monolayer ice

Guided by the experimental observations, we propose a monolayer ice structure for the 2×2 phase, which was obtained by full structural relaxation (Figs. 3c and 3d). In this model, water molecules arrange into a honeycomb lattice in which one molecule in each unit cell has one hydrogen atom oriented out of plane (upward), while the other one lies approximately in plane. This geometry yields an oxygen density ~2.67 times lower than that of interlocked bilayer ice, in excellent agreement with the coverage ratio of 2.74±0.18 measured in XPS. In addition, the calculated phonon spectrum of freestanding monolayer ice shows no imaginary frequencies, as shown in Supplementary Fig. S7, which confirmed the dynamic stability of freestanding monolayer ice.

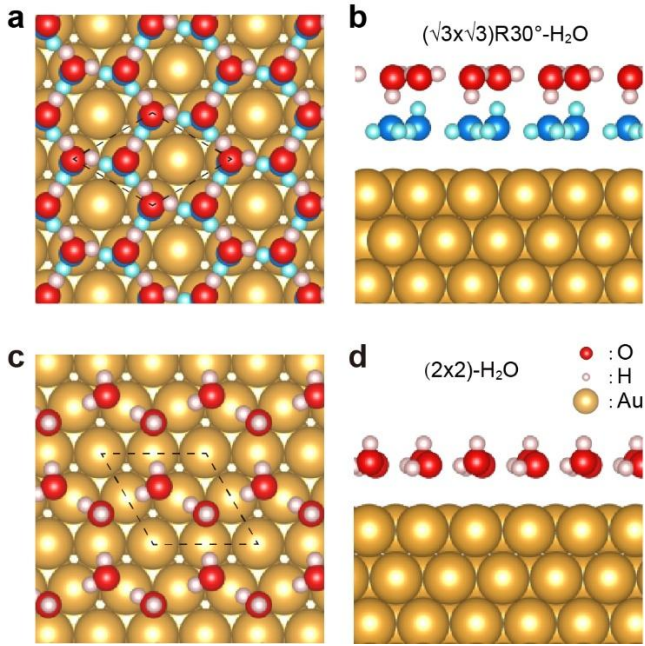

**Fig. 3.** Structure model of bilayer and monolayer ice on Au(111). (a, b) Top and side view of atomic structures of the bilayer ice that exhibit a (√3×√3)R30° superstructure on Au(111). (c, d) The same as (a) and (b) but for the monolayer ice that exhibits a 2×2 superstructure on Au(111).

ARPES measurements of monolayer and bilayer ice phases

With high-quality monolayer ice prepared, we investigated its electronic structure using ARPES. The pristine Au(111) surface exhibits its characteristic Shockley surface state (Figs. 4a and 4b). After adsorption of interlocked bilayer ice, no new dispersive states appear; instead, a non-dispersive feature emerges at approximately –2.7 eV below the Fermi level across the entire Brillouin zone (Figs. 4e and 4f). This flat band corresponds to the highest occupied molecular orbital (HOMO) of water on Au(111), consistent with earlier reports for water monomers on NaCl(001)/Au(111)(36).

Compared with the bilayer ice phase, no additional electronic states occur in the monolayer phase (Figs. 4i and 4j). The –2.7 eV flat band persists in the monolayer phase but with noticeably reduced spectral intensity (Figs. 4g and 4k), which is attributed to its lower molecular density relative to the bilayer structure. The absence of dispersive features in both phases indicates that the water molecules interact predominantly through localized hydrogen bonds, without significant orbital hybridization into extended covalent bands.

First-principles calculations for the bilayer structure reproduce the flat spectral weight at –2.7 eV in the calculated band structure and projected density of states (PDOS) along the M–Γ–M direction of Au(111) (Fig. 4h), while the calculations for the monolayer structure (Fig. 4l) also yields a flat band at the same binding energy, with reduced spectral weight, consistent with the ARPES observations.

Discussion

Our recent study demonstrated that an ordered $H_2O$–OH monolayer can form on Ag(111) via the LEEA growth method, indicating that low-energy electrons can effectively induce partial dissociation of $H_2O$ molecules on Ag(111). However, our experiments show no signatures of $H_2O$ dissociation on Au(111) under similar conditions.

We first provide a phenomenological picture of this discrepancy. The deeper 5d-band center of Au lowers the energy of the antibonding metal–adsorbate states and increases their effective filling, causing the adsorbate–metal interaction to be dominated by Pauli repulsion(37). Consequently, intact $H_2O$ molecules on Au(111) are more susceptible to electron-stimulated desorption under electron irradiation. In contrast, Ag(111) has a lower conduction-electron density and significantly weaker many-body screening. Injected electrons on $H_2O$ therefore relax more slowly into the Ag(111) substrate, allowing transient $H_2O^-$ resonances to persist long enough to break the O–H bond. Furthermore, the lower work function of Ag(111) places the unoccupied O–H antibonding state closer to the Fermi level, making it more accessible to electron population. Collectively, these factors render electron-induced $H_2O$ dissociation much more efficient on Ag(111) than on Au(111).

To further understand this discrepancy, we performed first-principles calculations of water adsorption and dissociation on both Au(111) and Ag(111) surfaces. The effect of electron beam irradiation on a 2D ice layer can manifest in two main ways: (i) desorption of intact $H_2O$ molecules, and (ii) dissociation into OH and H species. The adsorption energy per $H_2O$ molecule is defined as

$$E_{ad} = \frac{E_{tot} - E_{sub} - N\mu_{H_2O}}{N}$$

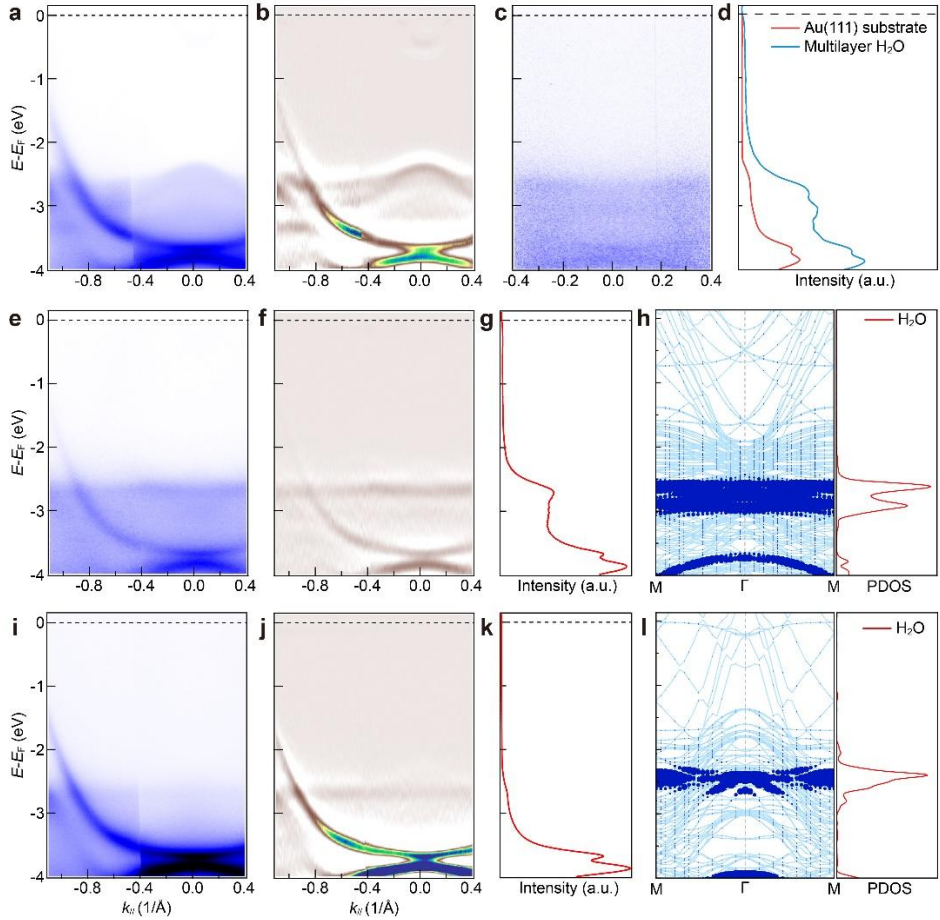

**Fig. 4.** (a) ARPES intensity map and (b) corresponding second-derivative image of the Au(111) surface along Γ–M. (c) ARPES intensity map of multilayer $H_2O$ on Au(111) without electron-beam irradiation. (d) EDC comparison between pristine Au(111) (red) and multilayer $H_2O$ (blue). (e, f) ARPES intensity map and second-derivative image of the (√3×√3)R30° bilayer ice phase. (g) EDC of the (√3×√3)R30° bilayer ice. (h) Calculated band structure (left) and projected density of states (right) for the (√3×√3)R30° bilayer ice. (i–l) Same measurements and calculations as in panels (e–h), but for the 2×2 monolayer ice phase.

where $E_{tot}$ and $E_{sub}$ are the total energies of the adsorption system and the substrate, respectively, $\mu_{H_2O}$ is the chemical potential of gas-phase $H_2O$, and $N$ is the number of water molecules.

The reaction energy for dissociation is given by

$$\Delta E = E_{diss} - E_{tot}$$

where $E_{diss}$ is the total energy of the dissociatively adsorbed final state. The activation energy is

$$E_{act} = E_{ts} - E_{tot}$$

with $E_{ts}$ being the transition-state energy along the minimum-energy pathway between intact and dissociated states.

The calculated adsorption, activation, and reaction energies for $H_2O$ on Au(111) and Ag(111) are summarized in Table 1. The adsorption energy of $H_2O$ on Ag(111) (−0.32 eV) is more negative than that on Au(111) (−0.3 eV), indicating stronger binding to Ag(111) and correspondingly weaker interaction with Au(111). As a result, water molecules are more readily desorbed from Au(111). In addition, both the activation barrier and the reaction energy for water dissociation are larger on Au(111) than on Ag(111), indicating that electron-induced dissociation of $H_2O$ is energetically less favorable on Au(111). Dissociation pathways of $H_2O$ on Au(111) and Ag(111) are shown in Supplementary Fig. S8.

Therefore, under electron beam irradiation alone, $H_2O$ molecules on Au(111) preferentially desorb rather than dissociate, explaining the absence of $H_2O$–OH formation on Au(111). Previous work has shown that water dissociation on Au surfaces can be achieved by lowering the dehydrogenation barrier, for example via pre-adsorbed oxygen(38-42). In the absence of such promoters, Au(111) remains resistant to electron-induced dissociation, in contrast to Ag(111). Controlled desorption of $H_2O$ molecules from the bilayer phase yields a metastable monolayer ice phase on Au(111).

To further validate the proposed structural model, we performed an extensive structural search starting from multiple distinct initial geometries. All tested configurations converged to only two stable structures: the H-up and H-down models, in which the hydrogen atoms of the water molecules point predominantly away from or toward the substrate, respectively. After full structural relax-ation, both models retain the characteristic honeycomb hydrogen-bond network. Adsorption energy calculations reveal that the two configurations are nearly degenerate in stability, with values of -0.343 eV/$H_2O$ for H-up and –0.372 eV/$H_2O$ for H-down. These adsorption energies are notably less negative than that of the interlocked bilayer ice on Au(111) (–0.544 eV/$H_2O$), indicating weaker binding in the monolayer phase. Furthermore, the calculated electronic structure of the H-up model shows better agreement with our experimental ARPES results, as the water-derived flat band appears at a binding energy closer to the observed value (Supplementary Fig. S9). Therefore, the H-up configuration is likely dominant in our experiments, although a minor contribution from H-down domains cannot be completely excluded (Supplementary Fig. S10 and Table S1). We note that the near degeneracy of the H-up and H-down configurations suggests the possibility of dynamic interconversion, potentially involving proton rearrangement, which may motivate future studies of proton dynamics in this system.

**Table 1.** Calculated adsorption, activation, and reaction energies for water adsorption and dissociation on Au(111) and Ag(111) surfaces, obtained without (PBE) and with (PBE+DFT-D3) dispersion corrections. A 3×3 surface supercell with a seven-layer slab was used in all simulations.

| Energy (eV) | Au (111) | | Ag (111) | |
|---|---|---|---|---|
| | PBE | PBE+DFT-D3 | PBE | PBE+DFT-D3 |
| $\Delta E_{ad}$ | -0.09 | -0.30 | -0.12 | -0.32 |
| $E_{act}$ | 3.26 | 2.90 | 2.69 | 2.63 |
| $\Delta E$ | 1.94 | 1.83 | 1.82 | 1.54 |

## Summary and Outlook

The realization of a monolayer ice phase on a hydrophobic metal surface marks a significant advance in the study of interfacial water. Beyond expanding the known phase diagram of 2D ice, this work demonstrates that targeted electron injection can serve as a versatile tool for engineering water structures at the molecular scale, offering precise control over hydrogen-bond networks without chemical modification. Such control may enable systematic exploration of low-dimensional quantum phenomena in light-element systems, including proton dynamics, confined hydrogen bonding, and cooperative dipole ordering. Moreover, the ability to access and stabilize unconventional ice phases on otherwise inert substrates could have broad implications for surface chemistry, catalysis, and nanoscale energy transfer at aqueous interfaces, ultimately bridging the gap between fundamental water science and functional device applications.

## Methods

### Sample preparation

The Au(111) single crystal was cleaned by repeated cycles of Ar+ ion sputtering and annealing. Ultrapure $H_2O$ (Alfa Aesar) was further purified under vacuum by multiple freeze–pump–thaw cycles to remove residual gaseous impurities. Water dosing onto Au(111) was carried out at 120 K for 2 min. All experiments, including ARPES, XPS, and LEED, were performed in situ in the same ultrahigh-vacuum chamber with a base pressure of ~1 × $10^{-10}$ mbar.

### LEED measurements and the LEEA growth method

LEED measurements were carried out *in situ* in the $H_2O$ growth chamber equipped with a SPECS ErLEED 150 reverse-view optics system. During data acquisition, the sample temperature was maintained at 120 K. The LEED patterns were recorded using an incident electron energy of 80 eV, a cathode current of 2.3 A, and a screen voltage of 6 kV. For LEEA growth, electrons were injected by increasing the beam energy to 280 eV. At each beam position, the sample was exposed until the diffraction pattern transitioned completely from the (√3×√3)R30° superstructure to the 2×2 superstructure. To prepare samples suitable for ARPES, the manipulator was raster-scanned across the entire surface to achieve a spatially uniform 2×2 phase.

### ARPES measurements

ARPES measurements were performed using a laboratory-based setup equipped with a SPECS PHOIBOS 150 electron energy analyser and a helium discharge lamp providing He Iα radiation (hν = 21.2 eV). During ARPES data acquisition, the sample temperature was maintained at ~70 K.

### XPS measurements

XPS measurements were carried out in the same chamber used for sample growth and ARPES. The experiments employed a SPECS XR 50 X-ray source with an aluminium anode (Al Kα, hν = 1486.6 eV). The sample temperature was maintained at 70 K during data acquisition. Binding energies were calibrated against the Fermi edge of a clean Au foil.

### First-principles calculations

The structural relaxations and electronic property calculations were performed within the framework of density functional theory (DFT)(43) with the projector-augmented wave (PAW) method(44), as implemented in the Vienna ab initio Simulation Pack (VASP) (45,46). The exchange–correlation interactions were treated within the generalized gradient approximation (GGA) of Perdew, Burke, and Ernzerhof (PBE)(47)A kinetic energy cutoff of 550 eV was employed for the plane-wave basis set. The Brillouin zone was sampled using a Γ-centered Monkhorst–Pack grid of 4×4×1(48). Structural optimizations were performed until the total energy and atomic forces converged to within $10^{-5}$ eV and 0.02 eV Å$^{-1}$, respectively. The Au(111) surface was modeled by a seven-layer slab. For the ice/Au systems, the in-plane lattice constants of Au were fixed, while all atomic coordinates were fully relaxed. To include the van der Waals interaction, DFT-D3 dispersion corrections were included(49). Phonon spectra were calculated using the finite displacement method implemented in the PHONOPY code(50) to determine the lattice dynamical stability of the ice monolayer.

## ASSOCIATED CONTENT

**Supporting Information.**
The Supporting Information is available free of charge at https://pubs.acs.org.

Growth and phase transition of two-dimensional ice on Au(111), Phonon spectrum of freestanding monolayer ice, dissociation pathways of $H_2O$ on Au(111) and Ag(111), DFT-calculated adsorption energies of 2D ice on Au(111) and electronic structure of monolayer ice on Au(111).

## AUTHOR INFORMATION

### Corresponding Author

**Yanchao Wang**–Key Laboratory of Material Simulation Methods & Software of Ministry of Education, College of Physics, Jilin University, Changchun 130012, China; Email: wyc@calypso.cn

**Lan Chen**–Institute of Physics, Chinese Academy of Sciences, Beijing 100190, China; School of Physical Sciences, University of Chinese Academy of Sciences, Beijing 100049, China; Email: lchen@iphy.ac.cn

**Baojie Feng**–Institute of Physics, Chinese Academy of Sciences, Beijing 100190, China; School of Physical Sciences, University of Chinese Academy of Sciences, Beijing 100049; Email: bjfeng@iphy.ac.cn

### Author Contributions

‡ Q.Z and M.X. contribute to this work.

### Notes

The authors declare no competing interest.

## ACKNOWLEDGMENT

This work was supported by the National Key R&D Program of China (Grants Nos. 2024YFA1408700, 2024YFA1408400, and 2022YFA1402304), the National Natural Science Foundation of China (Grants Nos. W2411004, 12374197, T2325028, 12134019, T2225013, and 12374010), the Beijing Natural Science Foundation (Grant No. JQ23001), and the CAS Project for Young Scientists in Basic Research (Grants No. YSBR-047).